%% file: paper.tex
\documentclass{article}
\pdfoutput=1
\usepackage{ijcai17}
\usepackage{times}
\usepackage{graphicx}
\usepackage{color}
\usepackage{threeparttable}
\usepackage{multirow}
\title{DeepFM: A Factorization-Machine based Neural Network for CTR Prediction}

\author{Huifeng Guo\thanks{This work is done when Huifeng Guo worked as intern at Noah's Ark Research Lab, Huawei.}$^1$ , Ruiming Tang$^2$, Yunming Ye\thanks{Corresponding Author.}$^1$, Zhenguo Li$^2$, Xiuqiang He$^2$\\
$^1$Shenzhen Graduate School, Harbin Institute of Technology, China\\
$^2$Noah's Ark Research Lab, Huawei, China\\
$^1$huifengguo@yeah.net, yeyunming@hit.edu.cn\\
$^2$\{tangruiming, li.zhenguo, hexiuqiang\}@huawei.com
}

\begin{document}

\maketitle

\begin{abstract}
Learning sophisticated feature interactions behind user behaviors is critical in maximizing CTR for recommender systems.
%For instance, as we observe in a mainstream apps market, people tend to download apps for food delivery at meal-time, which indicates that the interaction of app category and time-stamp is highly predictive of CTR.
%In general, CTR can be affected by profound feature interactions that are hard to engineering and have to be learned automatically from data.
Despite great progress, existing methods seem to have a strong bias towards low- or high-order interactions, or require expertise feature engineering. In this paper, we show that it is possible to derive an end-to-end learning model that emphasizes both low- and high-order feature interactions. The proposed model, DeepFM, combines the power of factorization machines for recommendation and deep learning for feature learning in a new neural network architecture. Compared to the latest Wide \& Deep model from Google, DeepFM has a shared input to its ``wide'' and ``deep'' parts, with no need of feature engineering besides raw features. Comprehensive experiments are conducted to demonstrate the effectiveness and efficiency of DeepFM over the existing models for CTR prediction, on both benchmark data and commercial data.

\end{abstract}

\input{intro.tex}

\input{models.tex}

\input{experiments.tex}

\input{related.tex}

\input{conclusion.tex}

%\newpage
%% The file named.bst is a bibliography style file for BibTeX 0.99c
\bibliographystyle{named}
\bibliography{complete}

\end{document}

%% file: intro.tex
\section{Introduction}\label{section:intro}
The prediction of click-through rate (CTR) is critical in recommender system, where the task is to estimate the probability a user will click on a recommended item. In many recommender systems the goal is to maximize the number of clicks, and so the items returned to a user can be ranked by estimated CTR; while in other application scenarios such as online advertising it is also important to improve revenue, and so the ranking strategy can be adjusted as CTR$\times$bid across all candidates, where ``bid'' is the benefit the system receives if the item is clicked by a user. In either case, it is clear that the key is in estimating CTR correctly.

%and it is found quite beneficial to model the sophisticated feature interactions behind user behaviors (e.g., \cite{fm,pnn,wide-n-deep})
\begin{figure}[ht]
\setlength{\abovecaptionskip}{0pt}%
\setlength{\belowcaptionskip}{-10pt}
\centering
\includegraphics[width=0.45\textwidth]{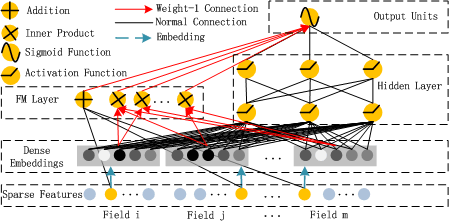}
\caption{\footnotesize{Wide \& deep architecture of DeepFM. The wide and deep component share the same input raw feature vector, which enables DeepFM to learn low- and high-order feature interactions simultaneously from the input raw features.}}\label{fig:architecture}
\end{figure}
It is important for CTR prediction to learn implicit feature interactions behind user click behaviors. By our study in a mainstream apps market, we found that people often download apps for food delivery at meal-time, suggesting that the (order-2) interaction between app category and time-stamp can be used as a signal for CTR. As a second observation, male teenagers like shooting games and RPG games, which means that the (order-3) interaction of app category, user gender and age is another signal for CTR. In general, such interactions of features behind user click behaviors can be highly sophisticated, where both low- and high-order feature interactions should play important roles. According to the insights of the Wide \& Deep model~\cite{wide-n-deep} from google, considering low- and high-order feature interactions \emph{simultaneously} brings additional improvement over the cases of considering either alone.

The key challenge is in effectively modeling feature interactions. Some feature interactions can be easily understood, thus can be designed by experts (like the instances above). However, most other feature interactions are hidden in data and difficult to identify \emph{a priori} (for instance, the classic association rule ``diaper and beer'' is mined from data, instead of discovering by experts), which can only be captured \emph{automatically} by machine learning. Even for easy-to-understand interactions, it seems unlikely for experts to model them exhaustively, especially when the number of features is large.

%Below we review previous efforts in learning feature interaction and then summarize our contributions.
%%\subsection{Background}\label{section:intro:back}
%Existing methods can be divided into three categories: ``wide'' models, which capture low-order feature interactions; deep models, which capture high-order feature interactions; and the latest Wide \& Deep models, which capture both low- and high-order feature interactions.
%

%\subsubsection{Wide models}
Despite their simplicity, generalized linear models, such as \emph{FTRL}~\cite{FTRL}, have shown decent performance in practice. However, a linear model lacks the ability to learn feature interactions, and a common practice is to manually include pairwise feature interactions in its feature vector. Such a method is hard to generalize to model high-order feature interactions or those never or rarely appear in the training data~\cite{fm}. \emph{Factorization Machines (FM)}~\cite{fm} model pairwise feature interactions as inner product of latent vectors between features and show very promising results. While in principle FM can model high-order feature interaction, in practice usually only order-2 feature interactions are considered due to high complexity.

%GBDT~\cite{GBDT}, random forest~\cite{RF} are two different Tree ensemble models. They are also proposed to explore unseen feature interactions, however, these models are still unable to fully explore useful feature patterns due to the limit of the model expressiveness.

%\subsubsection{Deep models}
As a powerful approach to learning feature representation, deep neural networks  have the potential to learn sophisticated feature interactions. Some ideas extend CNN and RNN for CTR predition \cite{cnn,rnn}, but CNN-based models are biased to the interactions between neighboring features while RNN-based models are more suitable for click data with sequential dependency. \cite{fnn} studies feature representations and proposes \emph{Factorization-machine supported Neural Network (FNN)}. This model pre-trains FM before applying DNN, thus limited by the capability of FM. Feature interaction is studied in \cite{pnn}, by introducing a product layer between embedding layer and fully-connected layer, and proposing the \emph{Product-based Neural Network} (\emph{PNN}). As noted in~\cite{wide-n-deep}, PNN and FNN, like other deep models, capture little low-order feature interactions, which are also essential for CTR prediction. To model both low- and high-order feature interactions, \cite{wide-n-deep} proposes an interesting hybrid network structure (\emph{Wide \& Deep}) that combines a linear (``wide'') model and a deep model. In this model, two different inputs are required for the ``wide part'' and ``deep part'', respectively, and the input of ``wide part" still relies on expertise feature engineering.

One can see that existing models are biased to low- or high-order feature interaction, or rely on feature engineering. In this paper, we show it is possible to derive a learning model that is able to learn feature interactions of all orders in an end-to-end manner, without any feature engineering besides raw features. Our main contributions are summarized as follows:
\begin{itemize}
\item We propose a new neural network model DeepFM (Figure~\ref{fig:architecture}) that integrates the architectures of FM and deep neural networks (DNN). It models low-order feature interactions like FM and models high-order feature interactions like DNN. Unlike the wide \& deep model~\cite{wide-n-deep}, DeepFM can be trained end-to-end without any feature engineering.
\item DeepFM can be trained efficiently because its wide part and deep part, unlike~\cite{wide-n-deep}, share the same input and also the embedding vector. In ~\cite{wide-n-deep}, the input vector can be of huge size as it includes manually designed pairwise feature interactions in the input vector of its wide part, which also greatly increases its complexity.
\item We evaluate DeepFM on both benchmark data and commercial data, which shows consistent improvement over existing models for CTR prediction.
\end{itemize}

%% file: models.tex
\section{Our Approach}\label{section:App}

%\subsection{Problem Definition}\label{section:App:pro}

Suppose the data set for training consists of $n$ instances $(\chi,y)$, where $\chi$ is an $m$-fields data record usually involving a pair of user and item, and $y\in\{0,1\}$ is the associated label indicating user click behaviors ($y=1$ means the user clicked the item, and $y=0$ otherwise). $\chi$ may include categorical fields (e.g., gender, location) and continuous fields (e.g., age). Each categorical field is represented as a vector of one-hot encoding, and each continuous field is represented as the value itself, or a vector of one-hot encoding after discretization. Then, each instance is converted to $(x,y)$ where $x=[x_{field_1},x_{field_2}, ...,x_{filed_j},...,x_{field_m}]$ is a $d$-dimensional vector, with $x_{field_j}$ being the vector representation of the $j$-th field of $\chi$. Normally, $x$ is high-dimensional and extremely sparse. The task of CTR prediction is to build a prediction model $\hat{y}=CTR\_model(x)$ to estimate the probability of a user clicking a specific app in a given context.

\subsection{DeepFM}\label{section:App:model}
%\begin{figure*}[!ht]
%\setlength{\abovecaptionskip}{0pt}%
%\setlength{\belowcaptionskip}{-10pt}
%\centering
%\includegraphics[width=0.96\textwidth]{img/architecture-wide.png}
%\caption{\footnotesize{The wide \& deep architecture of the proposed DeepFM. Note its wide (FM) component and deep component share the same input raw feature vector, and share the same embedding layer.}}\label{fig:architecture}
%\end{figure*}

We aim to learn both low- and high-order feature interactions. To this end, we propose a Factorization-Machine based neural network (DeepFM). As depicted in Figure~\ref{fig:architecture}\footnote{In all Figures of this paper, a \textbf{Normal Connection} in black refers to a connection with weight to be learned; a \textbf{Weight-1 Connection}, red arrow, is a connection with weight 1 by default; \textbf{Embedding}, blue dashed arrow, means a latent vector to be learned; \textbf{Addition} means adding all input together; \textbf{Product}, including \textbf{Inner-} and \textbf{Outer-Product}, means the output of this unit is the product of two input vector; \textbf{Sigmoid Function} is used as the output function in CTR prediction; \textbf{Activation Functions}, such as relu and tanh, are used for non-linearly transforming the signal.}, DeepFM consists of two components, \emph{FM component} and \emph{deep component}, that share  the same input. For feature $i$, a scalar $w_i$ is used to weigh its order-1 importance, a latent vector $V_i$ is used to measure its impact of interactions with other features. $V_i$ is fed in FM component to model order-2 feature interactions, and fed in deep component to model high-order feature interactions. All parameters, including $w_i$, $V_i$, and the network parameters ($W^{(l)}$, $b^{(l)}$ below) are trained jointly for the combined prediction model:
\begin{equation}
\hat{y}=sigmoid(y_{FM}+y_{DNN}),
\label{eq:FMNNCTR}
\end{equation}
where $\hat{y}\in (0,1)$ is the predicted CTR, $y_{FM}$ is the output of FM component, and $y_{DNN}$ is the output of deep component.
%\begin{equation}
%y_{DNN}=W^{|H|+1}\cdot a^{H}+b^{|H|+1}
%\label{eq:DNNoutput}
%\end{equation}
%where $|H|$ is the number of hidden layer.
%%\subsubsection{The FM Component}\label{section:App:model:fm}
%$y_{FM}$ is the output of FM component:
%\begin{equation}
%\centering
%y_{FM}=\left< w,x  \right> +\sum_{j_1=1}^{d}\sum_{j_2=j_1+1}^{d}\left< V_i,V_j  \right>  x_{j_1}\cdot x_{j_2}
%\label{eq:FM-model}
%\end{equation}
%$\hat{y}$ is the prediction, $x=[x_1,x_2,...,x_d]$ is a vector of $d$ features, $w=[w_1,w_2,...,w_d]$ are parameters to model the order-1 feature interactions, $V=[V_1,V_2,...,V_d]$ are parameters to model the order-2 feature interactions by inner product of the corresponding vectors, $V_i$ is the latent vector of feature $i$.
\subsubsection{FM Component}\label{section:App:model:fm}
\begin{figure}[ht]
\setlength{\abovecaptionskip}{0pt}%
\setlength{\belowcaptionskip}{-10pt}
\centering
\includegraphics[width=0.42\textwidth]{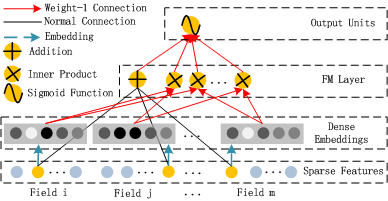}
\caption{\footnotesize{The architecture of FM. }}\label{fig:fm}
\end{figure}
The FM component is a factorization machine, which is proposed in \cite{fm} to learn feature interactions for recommendation. Besides a linear (order-1) interactions among features, FM models pairwise (order-2) feature interactions as inner product of respective feature latent vectors.
%Each latent vector is of length $k$, where k is a user-specified parameter.
It can capture order-2 feature interactions much more effectively than previous approaches especially when the dataset is sparse. In previous approaches, the parameter of an interaction of features $i$ and $j$ can be trained only when feature $i$ and feature $j$ both appear in the same data record. While in FM, it is measured via the inner product of their latent vectors $V_{i}$ and $V_{j}$. Thanks to this flexible design, FM can train latent vector $V_{i}$ ($V_{j}$) whenever $i$ (or $j$) appears in a data record. Therefore, feature interactions, which are never or rarely appeared in the training data, are better learnt by FM.

As Figure~\ref{fig:fm} shows, the output of FM is the summation of an \textbf{Addition} unit and a number of \textbf{Inner Product} units:
\begin{equation}
\centering
y_{FM}=\left< w,x  \right> +\sum_{j_1=1}^{d}\sum_{j_2=j_1+1}^{d}\left< V_i,V_j  \right>  x_{j_1}\cdot x_{j_2},
\label{eq:FM-model}
\end{equation}
where $w \in R^d$ and $V_i\in R^k$ ($k$ is given)\footnote{We omit a constant offset for simplicity.}. The Addition unit ($\left< w,x  \right>$) reflects the importance of order-1 features, and the Inner Product units represent the impact of order-2 feature interactions.

\subsubsection{Deep Component}\label{section:App:model:dnn}
\begin{figure}[ht]
\setlength{\abovecaptionskip}{0pt}%
\setlength{\belowcaptionskip}{-10pt}
\centering
\includegraphics[width=0.42\textwidth]{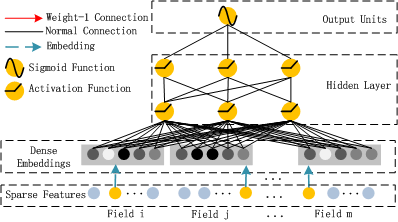}
\caption{\footnotesize{The architecture of DNN.}}\label{fig:dnn}
\end{figure}
The deep component is a feed-forward neural network, which is used to learn high-order feature interactions. As shown in Figure~\ref{fig:dnn}, a data record (a vector) is fed into the neural network. Compared to neural networks with image~\cite{residual2016} or audio~\cite{audioBoulanger-LewandowskiBV13} data as input, which is purely continuous and dense, the input of CTR prediction is quite different, which requires a new network architecture design. Specifically, the raw feature input vector for CTR prediction is usually highly sparse\footnote{Only one entry is non-zero for each field vector.}, super high-dimensional\footnote{E.g., in an app store of billion users, the one field vector for user ID is already of billion dimensions.}, categorical-continuous-mixed, and grouped in fields (e.g., gender, location, age). This suggests an embedding layer to compress the input vector to a low-dimensional, dense real-value vector before further feeding into the first hidden layer, otherwise the network can be overwhelming to train.
\begin{figure}[ht]
\setlength{\abovecaptionskip}{0pt}%
\setlength{\belowcaptionskip}{-10pt}
\begin{center}
\includegraphics[width=0.40\textwidth]{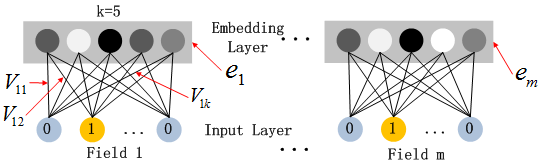}
\caption{\footnotesize{The structure of the embedding layer}}\label{fig:embed}
\end{center}
\end{figure}
%$\mathbf{V^T_{start_{field_1}:end_{field_1}}\cdot x_{start_{field_1}:end_{field_1}}}$
%$\mathbf{e_{n}}$$\mathbf{e_{1}}$

Figure \ref{fig:embed} highlights the sub-network structure from the input layer to the embedding layer. We would like to point out the two interesting features of this network structure: 1) while the lengths of different input field vectors can be different, their embeddings are of the same size ($k$); 2) the latent feature vectors ($V$) in FM now server as network weights which are learned and used to compress the input field vectors to the embedding vectors. In \cite{fnn}, $V$ is pre-trained by FM and used as initialization. In this work, rather than using the latent feature vectors of FM to initialize the networks as in \cite{fnn}, we include the FM model as part of our overall learning architecture, in addition to the other DNN model. As such, we eliminate the need of pre-training by FM and instead jointly train the overall network in an end-to-end manner.
Denote the output of the embedding layer as:
\begin{equation}
\label{eq:embed}
a^{(0)}=[e_{1},e_{2},...,e_{m}],
 \end{equation}
where $e_{i}$ is the embedding of $i$-th field and $m$ is the number of fields. Then, $a^{(0)}$ is fed into the deep neural network, and the forward process is:

%each field, e.g., \textbf{city}, have multiple units, each of which presents a specific value of this field, e.g., \textbf{city=London}. Embedding each field to a latent vector independently results in reducing the embedding parameters from ${\mid a^{(0)}\mid}\cdot {\mid x \mid }$ to $k\cdot {\mid x \mid }$, where k is a small user-specified parameter to define the dimension of latent vectors. The output of the embedding layer is:
%\begin{equation}
%\label{eq:embed}
%a^{(0)}=[e_{field\_1},e_{field\_2},...,e_{|filed|}]
% \end{equation}
%where $a^{0}$ is fed into the first hidden layer, and $e_{field\_i}$ is the embedding of $i$-th field:
%\begin{equation}
%e_{field\_i} = V[{start_{field\_i}:end_{field\_i}}]\cdot x_{field\_i}
%\label{eq:embedconcat}
%\end{equation}
%where start$_{field\_i}$ and end$_{filed\_i}$ are starting and ending feature indexes of the $i$-th field.

\begin{equation}
\centering
\label{eq:h2h}
a^{(l+1)} = \sigma(W^{(l)}a^{(l)} + b^{(l)}),
\end{equation}
where $l$ is the layer depth and $\sigma$ is an activation function. $a^{(l)}$, $W^{(l)}$, $b^{(l)}$ are the output, model weight, and bias of the $l$-th layer. After that, a dense real-value feature vector is generated, which is finally fed into the sigmoid function for CTR prediction: $y_{DNN}=\sigma({W^{|H|+1}\cdot a^{H}+b^{|H|+1}})$, where $|H|$ is the number of hidden layers.
%\begin{equation}
%y_{DNN}=W^{|H|+1}\cdot a^{H}+b^{|H|+1}
%\label{eq:DNNoutput}
%\end{equation}

It is worth pointing out that FM component and deep component share the same feature embedding, which brings two important benefits: 1) it learns both low- and high-order feature interactions from raw features; 2) there is no need for expertise feature engineering of the input, as required in Wide \& Deep~\cite{wide-n-deep}.

\subsection{Relationship with the other Neural Networks}\label{section:App:rela}

Inspired by the enormous success of deep learning in various applications, several deep models for CTR prediction are developed recently. This section compares the proposed DeepFM with existing deep models for CTR prediction.

\begin{figure*}[ht]
\setlength{\abovecaptionskip}{0pt}%
\setlength{\belowcaptionskip}{-10pt}
\centering
\includegraphics[width=0.92\textwidth]{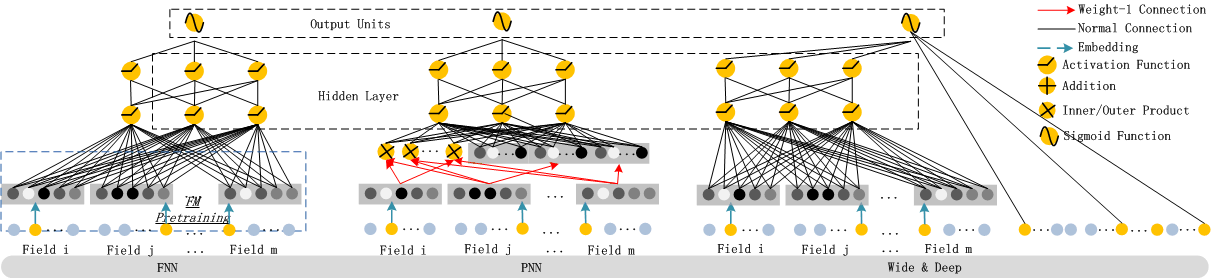}
\caption{\footnotesize{The architectures of existing deep models for CTR prediction: FNN, PNN, Wide \& Deep Model}}\label{fig:othermodel}
\end{figure*}

%\subsubsection{Factorization-machine supported Neural Network (FNN)}\label{section:App:rela:fnn}

\noindent\textbf{FNN:} As Figure~\ref{fig:othermodel} (left) shows, FNN is a FM-initialized feed-forward neural network~\cite{fnn}. The FM pre-training strategy results in two limitations: 1) the embedding parameters might be over affected by FM; 2) the efficiency is reduced by the overhead introduced by the pre-training stage. In addition, FNN captures only high-order feature interactions. In contrast, DeepFM needs no pre-training and learns both high- and low-order feature interactions.

%\subsubsection{Product-based Neural Network (PNN)}\label{section:App:rela:pnn}

\noindent\textbf{PNN:} For the purpose of capturing high-order feature interactions, PNN imposes a product layer between the embedding layer and the first hidden layer~\cite{pnn}. According to different types of product operation, there are three variants: IPNN, OPNN, and PNN$\ast$, where IPNN is based on inner product of vectors, OPNN is based on outer product, and PNN$\ast$ is based on both inner and outer products.

To make the computation more efficient, the authors proposed the approximated computations of both inner and outer products: 1) the inner product is approximately computed by eliminating some neurons; 2) the outer product is approximately computed by compressing $m$ $k$-dimensional feature vectors to one $k$-dimensional vector. However, we find that the outer product is less reliable than the inner product, since the approximated computation of outer product loses much information that makes the result unstable. Although inner product is more reliable, it still suffers from high computational complexity, because the output of the product layer is connected to all neurons of the first hidden layer. Different from PNN, the output of the product layer in DeepFM only connects to the final output layer (one neuron). Like FNN, all PNNs ignore low-order feature interactions.

%\subsubsection{Wide \& Deep}\label{section:App:rela:widedeep}

%\begin{figure}[ht]
%\centering
%\includegraphics[width=0.48\textwidth]{img/wide-deep.png}
%\caption{{The Wide \& Deep model}}\label{fig:widedeep}
%\end{figure}

\noindent\textbf{Wide \& Deep:} Wide \& Deep (Figure~\ref{fig:othermodel} (right)) is proposed by Google to model low- and high-order feature interactions simultaneously. As shown in~\cite{wide-n-deep}, there is a need for expertise feature engineering on the input to the ``wide" part (for instance, cross-product of users' install apps and impression apps in app recommendation). In contrast, DeepFM needs no such expertise knowledge to handle the input by learning directly from the input raw features.

A straightforward extension to this model is replacing LR by FM  (we also evaluate this extension in Section~\ref{section:exp}). This extension is similar to DeepFM, but DeepFM shares the feature embedding between the FM and deep component. The sharing strategy of feature embedding influences (in back-propagate manner) the feature representation by both low- and high-order feature interactions, which models the representation more precisely.

\noindent\textbf{Summarizations:} To summarize, the relationship between DeepFM and the other deep models in four aspects is presented in Table~\ref{table:Modelcompare}. As can be seen, DeepFM is the only model that requires no pre-training and no feature engineering, and captures both low- and high-order feature interactions.

\begin{table}
\centering
\scriptsize
\caption{\footnotesize{Comparison of deep models for CTR prediction}}\label{table:Modelcompare}
\begin{tabular}{|c|c|c|c|c|}
\hline
 &  No& High-order  & Low-order  & No Feature \\
 &Pre-training     & Features     & Features    & Engineering \\ \hline
FNN & $\times$ & $\surd$  & $\times$ & $\surd$ \\ \hline
PNN & ${\surd}$ & ${\surd}$  & ${\times}$ & ${\surd}$ \\ \hline
Wide \& Deep & ${\surd}$   & ${\surd}$  & ${\surd}$  & ${\times}$\\ \hline
DeepFM & ${\surd}$ & ${\surd}$  & ${\surd}$  & ${\surd}$ \\ \hline
\end{tabular}
\end{table}

%% file: experiments.tex
\section{Experiments}\label{section:exp}

In this section, we compare our proposed DeepFM and the other state-of-the-art models empirically. The evaluation result indicates that our proposed DeepFM is more effective than any other state-of-the-art model and the efficiency of DeepFM is comparable to the best ones among the others.

\subsection{Experiment Setup}\label{sec:exp:set}

\subsubsection{Datasets}\label{sec:exp:set:data}

We evaluate the effectiveness and efficiency of our proposed DeepFM on the following two datasets.

\noindent\textbf{1) Criteo Dataset}: Criteo dataset \footnote{http://labs.criteo.com/downloads/2014-kaggle-display-advertising-challenge-dataset/} includes 45 million users' click records. There are 13 continuous features and 26 categorical ones. We split the dataset randomly into two parts: 90\% is for training, while the rest 10\% is for testing.

\noindent\textbf{2) Company$\ast$ Dataset}: In order to verify the performance of DeepFM in real industrial CTR prediction, we conduct experiment on Company$\ast$ dataset. We collect 7 consecutive days of users' click records from the game center of the Company$\ast$ App Store for training, and the next 1 day for testing. There are around 1 billion records in the whole collected dataset. In this dataset, there are app features (e.g., identification, category, and etc), user features (e.g., user's downloaded apps, and etc), and context features (e.g., operation time, and etc).

\subsubsection{Evaluation Metrics}\label{sec:exp:set:metric}

We use two evaluation metrics in our experiments: \textbf{AUC} (Area Under ROC) and \textbf{Logloss} (cross entropy).

\subsubsection{Model Comparison}\label{sec:exp:set:comparemodel}

We compare 9 models in our experiments: \textbf{LR}, \textbf{FM}, \textbf{FNN}, \textbf{PNN (three variants)}, \textbf{Wide \& Deep}, and \textbf{DeepFM}. In the Wide \& Deep model, for the purpose of eliminating feature engineering effort, we also adapt the original Wide \& Deep model by replacing LR by FM as the wide part. In order to distinguish these two variants of Wide \& Deep, we name them LR \& DNN and FM \& DNN, respectively.\footnote{We do not use the Wide \& Deep API released by Google, as the efficiency of that implementation is very low. We implement Wide \& Deep by ourselves by simplifying it with shared optimizer for both deep and wide part.}

\subsubsection{Parameter Settings}\label{sec:exp:set:hyper}

To evaluate the models on Criteo dataset, we follow the parameter settings in \cite{pnn} for FNN and PNN: (1) dropout: 0.5; (2) network structure: 400-400-400; (3) optimizer: Adam; (4) activation function: tanh for IPNN, relu for other deep models. To be fair, our proposed DeepFM uses the same setting. The optimizers of LR and FM are FTRL and Adam respectively, and the latent dimension of FM is 10.

To achieve the best performance for each individual model on Company$\ast$ dataset, we conducted carefully parameter study, which is discussed in Section~\ref{sec:exp:hyper}.
%We present in Table~\ref{table:comhyper} the hyper-parameters of deep models that can achieve the best performance on Company$\ast$ dataset.
%
%\begin{table}[ht]
%
%\centering
%%\scriptsize
%\caption{Hyper-parameters of deep model on Company$\ast$}\label{table:comhyper}
%\begin{tabular}{|c|c|c|c|}
%\hline
%
%&Activate&&Model \\
%&Function&Dropout&Structure \\ \hline
%FNN  & relu & 0.8 & 150-300-150\\ \hline
%IPNN  & tanh & 0.9 & 400-400-400\\ \hline
%OPNN   & relu & 0.6 & 500-400-300\\ \hline
%PNN$\ast$  & relu & 0.8 & 100-200-300\\ \hline
%LR \& DNN   & relu & 0.7 & 400-400-400\\ \hline
%FM \& DNN   & relu & 0.8 & 900-800-700\\ \hline
%DeepFM  & relu & 0.9 & 400-400-400\\ \hline
%\end{tabular}
%\end{table}

\subsection{Performance Evaluation}\label{sec:exp:perfor}

In this section, we evaluate the models listed in Section~\ref{sec:exp:set:comparemodel} on the two datasets to compare their effectiveness and efficiency.

\subsubsection{Efficiency Comparison}\label{sec:exp:perfor:time}

\begin{figure}[ht]
\setlength{\abovecaptionskip}{0pt}%
\setlength{\belowcaptionskip}{-10pt}
\centering
\begin{minipage}[b]{0.5\textwidth}
\includegraphics[width=0.45\textwidth]{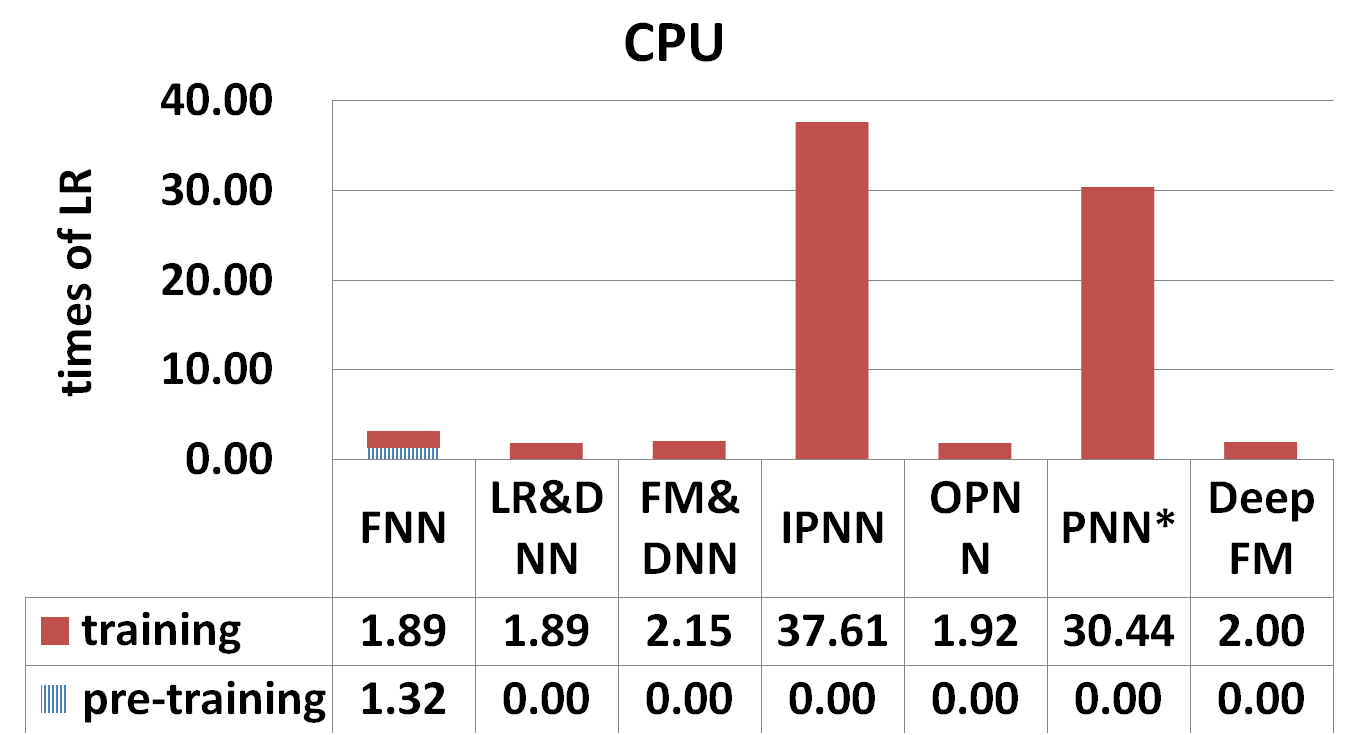}
\includegraphics[width=0.45\textwidth]{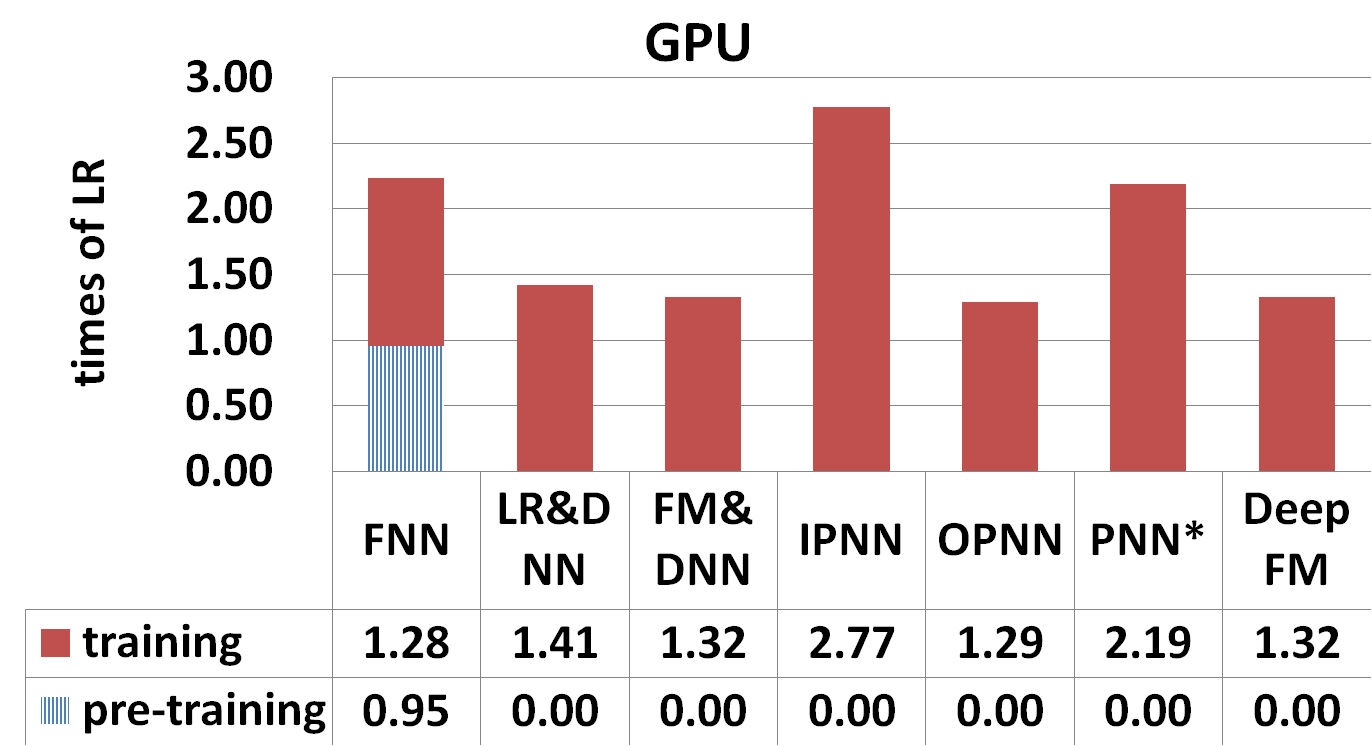}
\end{minipage}
\caption{\footnotesize{Time comparison.}}\label{fig:time}
\end{figure}

The efficiency of deep learning models is important to real-world applications. We compare the efficiency of different models on Criteo dataset by the following formula: $\frac{|training\ time\ of\ deep\ CTR\ model|}{|training\ time\ of\ LR|}$. The results are shown in Figure~\ref{fig:time}, including the tests on CPU (left) and GPU (right), where we have the following observations:
1) pre-training of FNN makes it less efficient;
2) Although the speed up of IPNN and PNN$\ast$ on GPU is higher than the other models, they are still computationally expensive because of the inefficient inner product operations;
3) The DeepFM achieves almost the most efficient in both tests.

%Test on CPU suggests that efficiency rank of the models is: LR \& DNN, OPNN, DeepFM, FM \& DNN, FNN, PNN$\ast$, IPNN. This rank is consistent with our theoretical analysis: pre-training of FNN makes it less efficient; IPNN and PNN$\ast$ are computationally expensive because of the inefficient inner product; OPNN is much faster than the IPNN and PNN$\ast$ because of computation approximation. Although DeepFM is only ranked the third place over all the 7 deep models, the difference between DeepFM and the winner LR \& DNN is minor.
%
%GPU is able to accelerate the matrix multiplication of inner product significantly, therefore the speed up of IPNN and PNN$\ast$ is higher than the other models. Even so, the efficiency order of the models is changed slightly: OPNN, DeepFM, FM \& DNN, LR \& DNN, PNN$\ast$, FNN, IPNN. As we can see, in GPU environment, the pre-training of FNN makes it even more unacceptable to use in real cases. The DeepFM achieves almost the best performance in terms of efficiency.

\subsubsection{Effectiveness Comparison}
The performance for CTR prediction of different models on Criteo dataset and Company$\ast$ dataset is shown in Table~\ref{table:performance}, where we have the following observations:
\begin{itemize}
\item Learning feature interactions improves the performance of CTR prediction model. This observation is from the fact that LR (which is the only model that does not consider feature interactions) performs worse than the other models. As the best model, DeepFM outperforms LR by 0.86\% and 4.18\% in terms of AUC (1.15\% and 5.60\% in terms of Logloss) on Company$\ast$ and Criteo datasets.
\item Learning high- and low-order feature interactions simultaneously and properly improves the performance of CTR prediction model. DeepFM outperforms the models that learn only low-order feature interactions (namely, FM) or high-order feature interactions (namely, FNN, IPNN, OPNN, PNN$\ast$). Compared to the second best model, DeepFM achieves more than 0.37\% and 0.25\% in terms of AUC (0.42\% and 0.29\% in terms of Logloss) on Company$\ast$ and Criteo datasets.
\item Learning high- and low-order feature interactions simultaneously while sharing the same feature embedding for high- and low-order feature interactions learning improves the performance of CTR prediction model. DeepFM outperforms the models that learn high- and low-order feature interactions using separate feature embeddings (namely, LR \& DNN and FM \& DNN). Compared to these two models, DeepFM achieves more than 0.48\% and 0.33\% in terms of AUC (0.61\% and 0.66\% in terms of Logloss) on Company$\ast$ and Criteo datasets.
\end{itemize}
%Table~\ref{table:performance} presents the overall performance of different models on Criteo dataset and Company$\ast$ dataset. As can be observed, DeepFM outperforms all the other models in terms of AUC and Logloss, on both datasets. This improvement empirically verifies our theoretic analysis when we design DeepFM in Section~\ref{section:App}. DeepFM achieves better performance than LR, FM, FNN and PNN, due to the fact that DeepFM considers both high- and low-order feature interactions when making prediction. Moreover, DeepFM outperforms LR\&DNN and FM\&DNN, because DeepFM shares the feature embedding between deep and FM component which makes the feature representation more precise.

\begin{table}[ht]
\centering
\footnotesize
\caption{\footnotesize{Performance on CTR prediction.}}\label{table:performance}
\begin{tabular}{|c|c|c|c|c|}
\hline
\multirow{2}{*}{} & \multicolumn{2}{c|}{Company$\ast$} & \multicolumn{2}{c|}{Criteo} \\
 \cline{2-5}
 & AUC & LogLoss & AUC & LogLoss\\ \hline
LR      & 0.8640    & 0.02648   & 0.7686 & 0.47762\\ \hline
FM      & 0.8678    & 0.02633   & 0.7892 & 0.46077\\ \hline
FNN     & 0.8683    & 0.02629   & 0.7963 & 0.45738\\ \hline
IPNN    & 0.8664    & 0.02637   & 0.7972 & 0.45323\\ \hline
OPNN    & 0.8658    & 0.02641   & 0.7982 & 0.45256\\ \hline
PNN$\ast$    & 0.8672    & 0.02636   & 0.7987 & 0.45214\\ \hline
LR \& DNN & 0.8673    & 0.02634   & 0.7981 & 0.46772\\ \hline
FM \& DNN & 0.8661    & 0.02640   & 0.7850 & 0.45382\\ \hline
DeepFM  & \textbf{0.8715}    &  \textbf{0.02618}  & \textbf{0.8007} & \textbf{0.45083}\\ \hline
\end{tabular}
\end{table}

%the performance of FM is better than LR. Specifically, it is 0.58\% lower in terms of LogLoss and 0.43\% higher in terms of AUC than LR on Company$\ast$, and 3.53\% lower in terms of LogLoss and 2.68\% higher in terms of AUC than LR on Criteo. This improvement comes from the ability of FM to capture order-2 feature interactions automatically. In addition, when considering both low- and high-order feature interactions, DeepFM obtains the highest performance. It is 1.15\% lower in terms of LogLoss and 0.86\% higher in terms of AUC than LR on Company$\ast$, and 5.60\% lower in terms of LogLoss and 4.18\% higher in terms of AUC than LR on Criteo.

Overall, our proposed DeepFM model beats the competitors by more than 0.37\% and 0.42\% in terms of AUC and Logloss on Company$\ast$ dataset, respectively. In fact, a small improvement in offline AUC evaluation is likely to lead to a significant increase in online CTR. As reported in ~\cite{wide-n-deep}, compared with LR, Wide \& Deep improves AUC by 0.275\% (offline) and the improvement of online CTR is 3.9\%. The daily turnover of Company$\ast$'s App Store is millions of dollars, therefore even several percents lift in CTR brings extra millions of dollars each year.

\subsection{Hyper-Parameter Study}\label{sec:exp:hyper}

We study the impact of different hyper-parameters of different deep models, on Company$\ast$ dataset. The order is: 1) activation functions; 2) dropout rate; 3) number of neurons per layer; 4) number of hidden layers; 5) network shape.

\subsubsection{Activation Function}\label{sec:exp:hyper:act}

\begin{figure}[ht]
\setlength{\abovecaptionskip}{0pt}%
\setlength{\belowcaptionskip}{-10pt}
\centering
\begin{minipage}[b]{0.5\textwidth}
\includegraphics[width=0.48\textwidth]{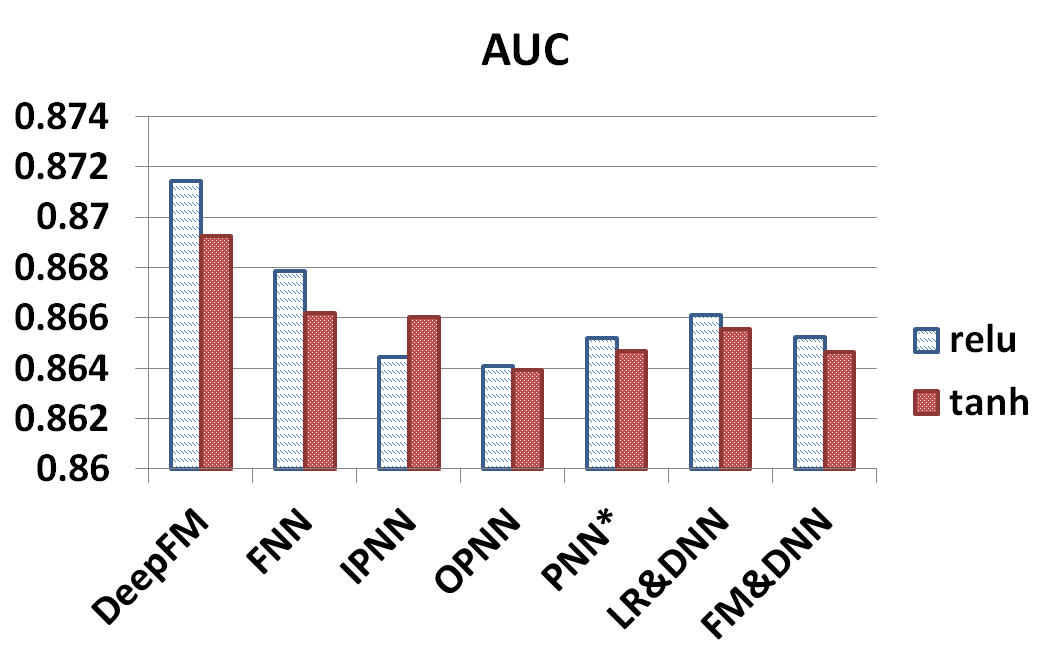}
\includegraphics[width=0.48\textwidth]{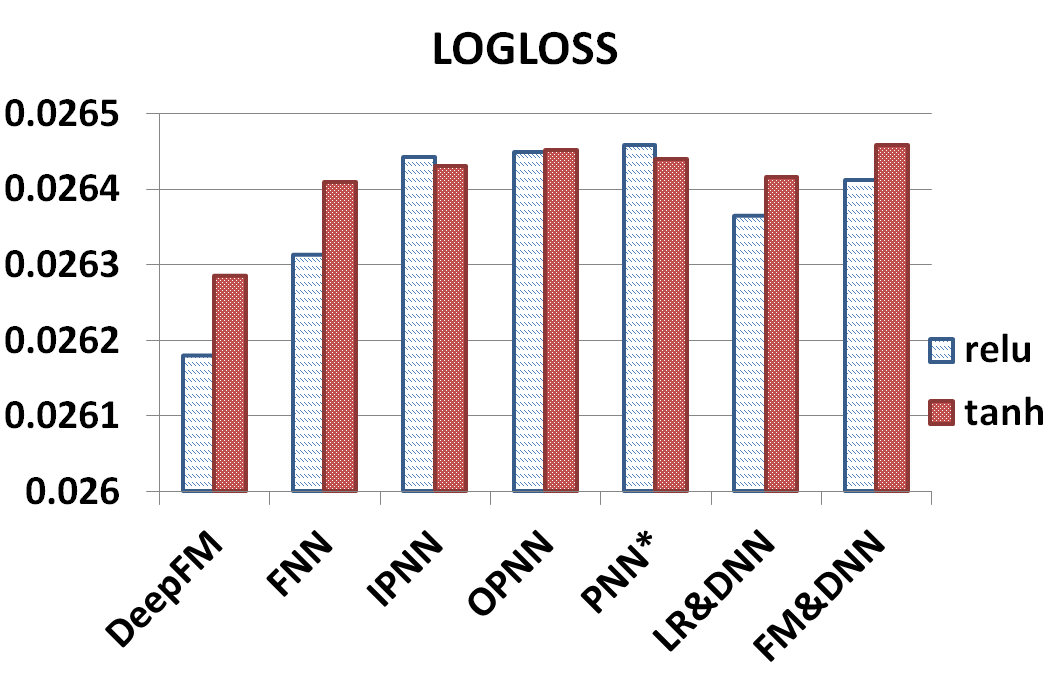}
\end{minipage}
\caption{\footnotesize{AUC and Logloss comparison of activation functions.}}\label{fig:act}
\end{figure}

According to \cite{pnn}, \emph{relu} and \emph{tanh} are more suitable for deep models than \emph{sigmoid}. In this paper, we compare the performance of deep models when applying \emph{relu} and \emph{tanh}. As shown in Figure~\ref{fig:act}, relu is more appropriate than tanh for all the deep models, except for IPNN. Possible reason is that relu induces sparsity.

\subsubsection{Dropout}\label{sec:exp:hyper:drop}

\begin{figure}[ht]
\setlength{\abovecaptionskip}{0pt}%
\setlength{\belowcaptionskip}{-10pt}
\centering
\begin{minipage}[b]{0.5\textwidth}
\includegraphics[width=0.48\textwidth]{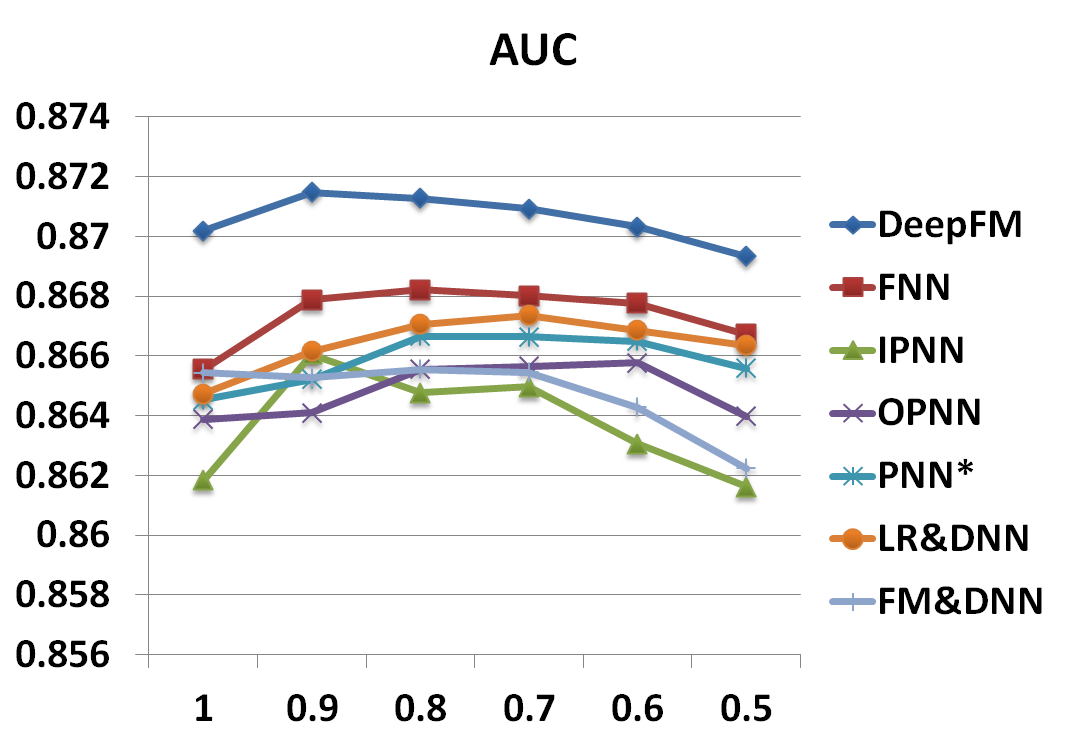}
\includegraphics[width=0.48\textwidth]{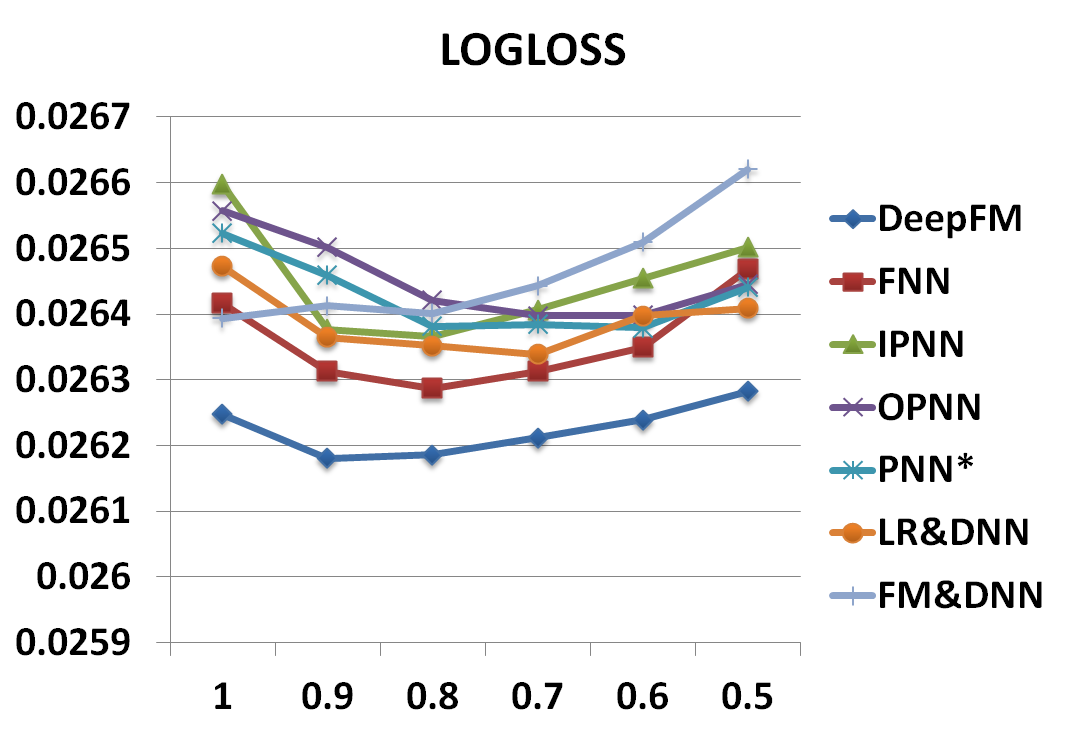}
\end{minipage}
\caption{\footnotesize{AUC and Logloss comparison of dropout.}}\label{fig:drop}
\end{figure}

Dropout~\cite{dropout14} refers to the probability that a neuron is kept in the network. Dropout is a regularization technique to compromise the precision and the complexity of the neural network. We set the dropout to be 1.0, 0.9, 0.8, 0.7, 0.6, 0.5. As shown in Figure~\ref{fig:drop}, all the models are able to reach their own best performance when the dropout is properly set (from 0.6 to 0.9). The result shows that adding reasonable randomness to model can strengthen model's robustness.

\subsubsection{Number of Neurons per Layer}\label{sec:exp:hyper:neuron}

\begin{figure}[ht]
\setlength{\abovecaptionskip}{0pt}%
\setlength{\belowcaptionskip}{-10pt}
\centering
\begin{minipage}[b]{0.5\textwidth}
\includegraphics[width=0.48\textwidth]{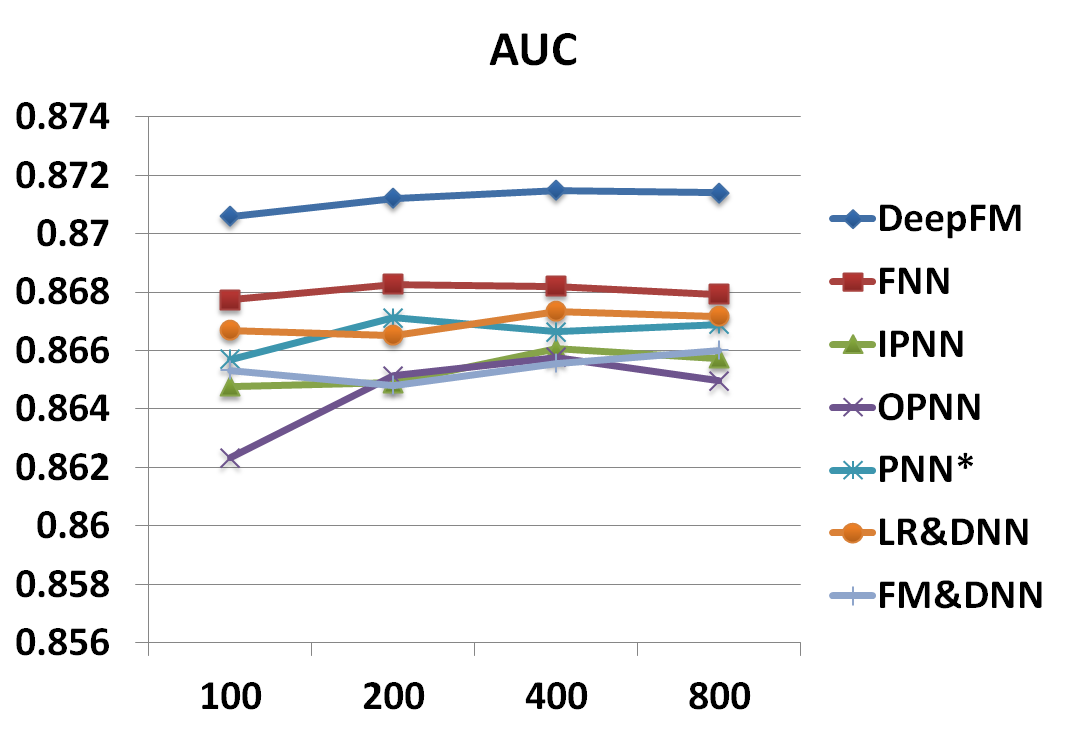}
\includegraphics[width=0.48\textwidth]{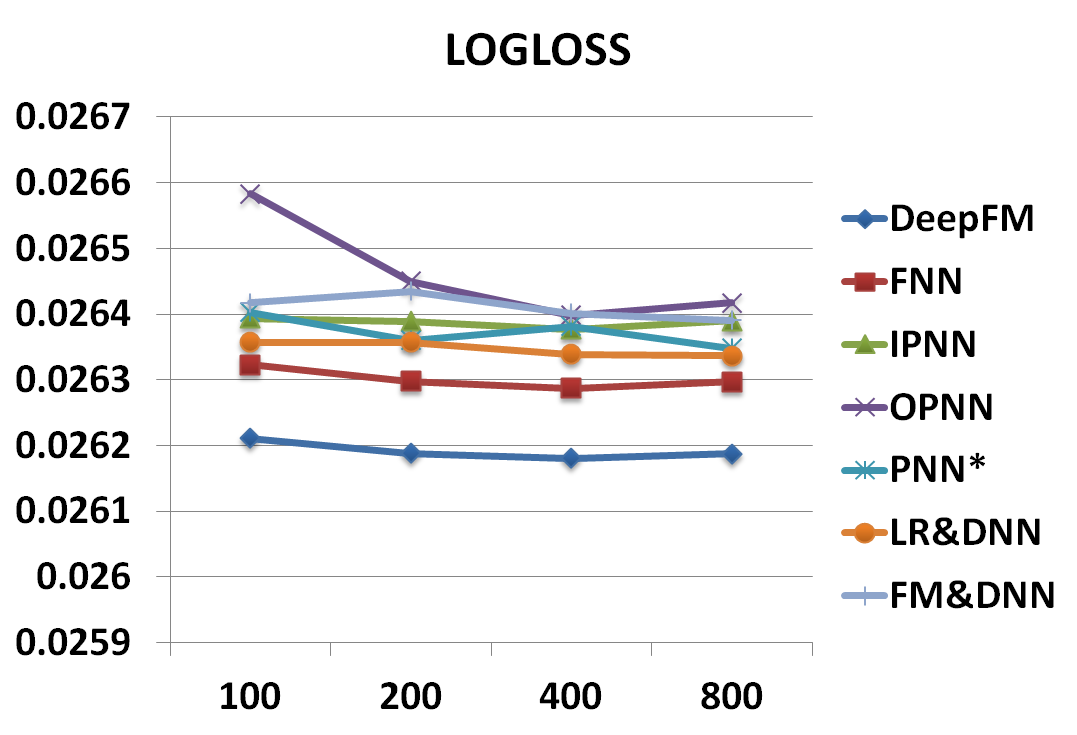}
\end{minipage}
\caption{\footnotesize{AUC and Logloss comparison of number of neurons.}}\label{fig:neuron}
\end{figure}

When other factors remain the same, increasing the number of neurons per layer introduces complexity. As we can observe from Figure~\ref{fig:neuron}, increasing the number of neurons does not always bring benefit. For instance, DeepFM performs stably when the number of neurons per layer is increased from 400 to 800; even worse, OPNN performs worse when we increase the number of neurons from 400 to 800. This is because an over-complicated model is easy to overfit. In our dataset, 200 or 400 neurons per layer is a good choice.

\subsubsection{Number of Hidden Layers}\label{sec:exp:hyper:layer}

\begin{figure}[ht]
\setlength{\abovecaptionskip}{0pt}%
\setlength{\belowcaptionskip}{-10pt}
\centering
\begin{minipage}[b]{0.5\textwidth}
\includegraphics[width=0.48\textwidth]{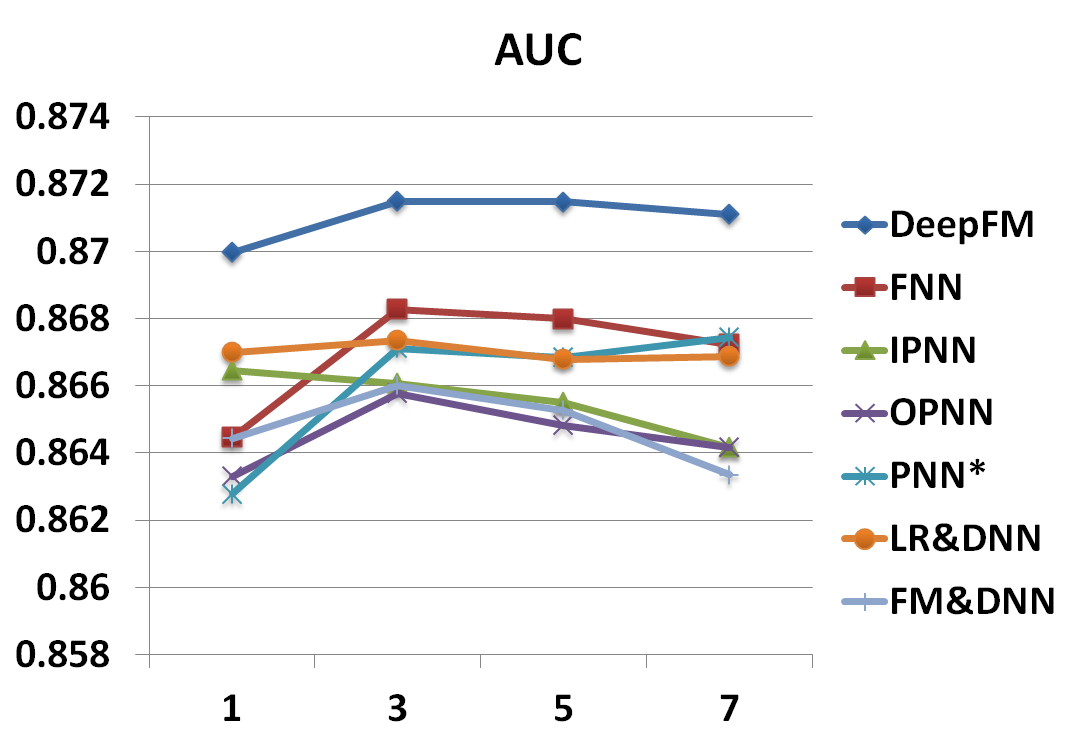}
\includegraphics[width=0.48\textwidth]{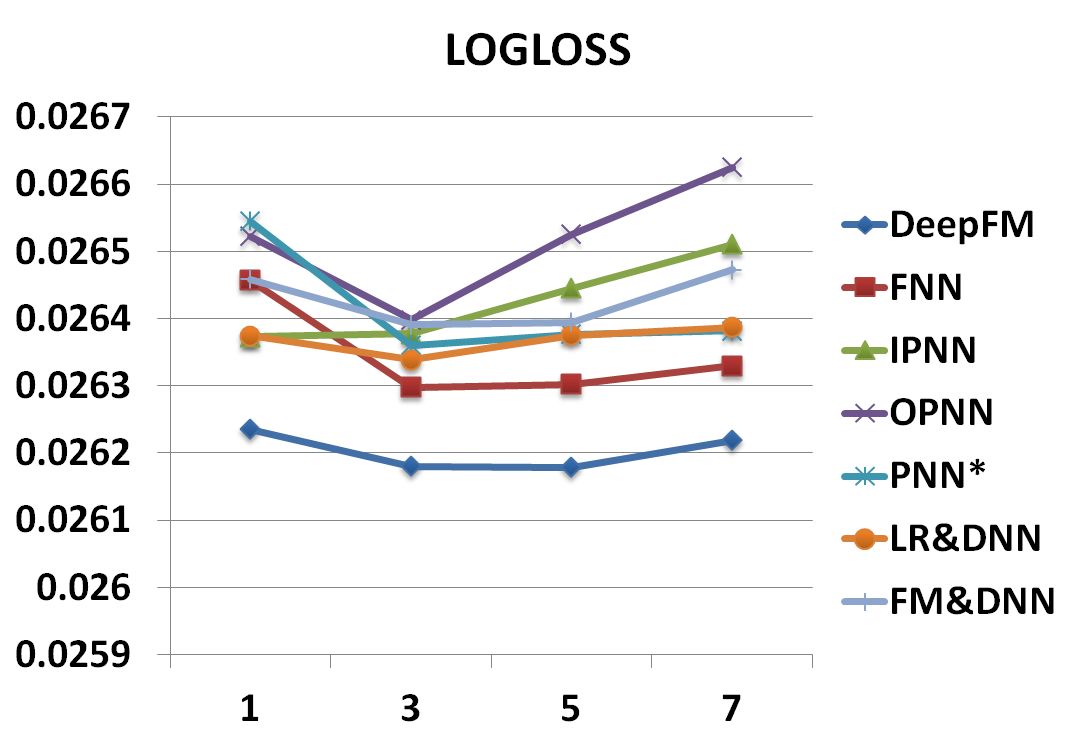}
\end{minipage}
\caption{\footnotesize{AUC and Logloss comparison of number of layers.}}\label{fig:layer}
\end{figure}

As presented in Figure~\ref{fig:layer}, increasing number of hidden layers improves the performance of the models at the beginning, however, their performance is degraded if the number of hidden layers keeps increasing. This phenomenon is also because of overfitting.

\subsubsection{Network Shape}\label{sec:exp:hyper:shape}

\begin{figure}[ht]
\setlength{\abovecaptionskip}{0pt}%
\setlength{\belowcaptionskip}{-10pt}
\centering
\begin{minipage}[b]{0.5\textwidth}
\includegraphics[width=0.48\textwidth]{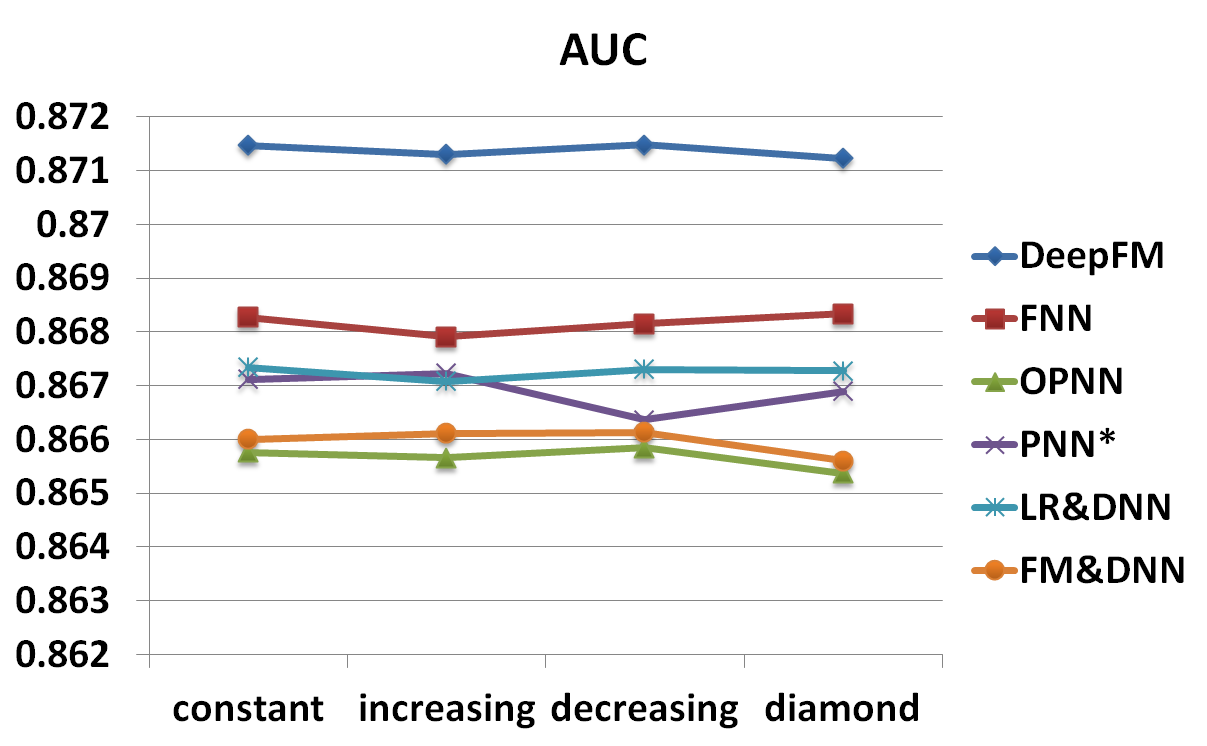}
\includegraphics[width=0.48\textwidth]{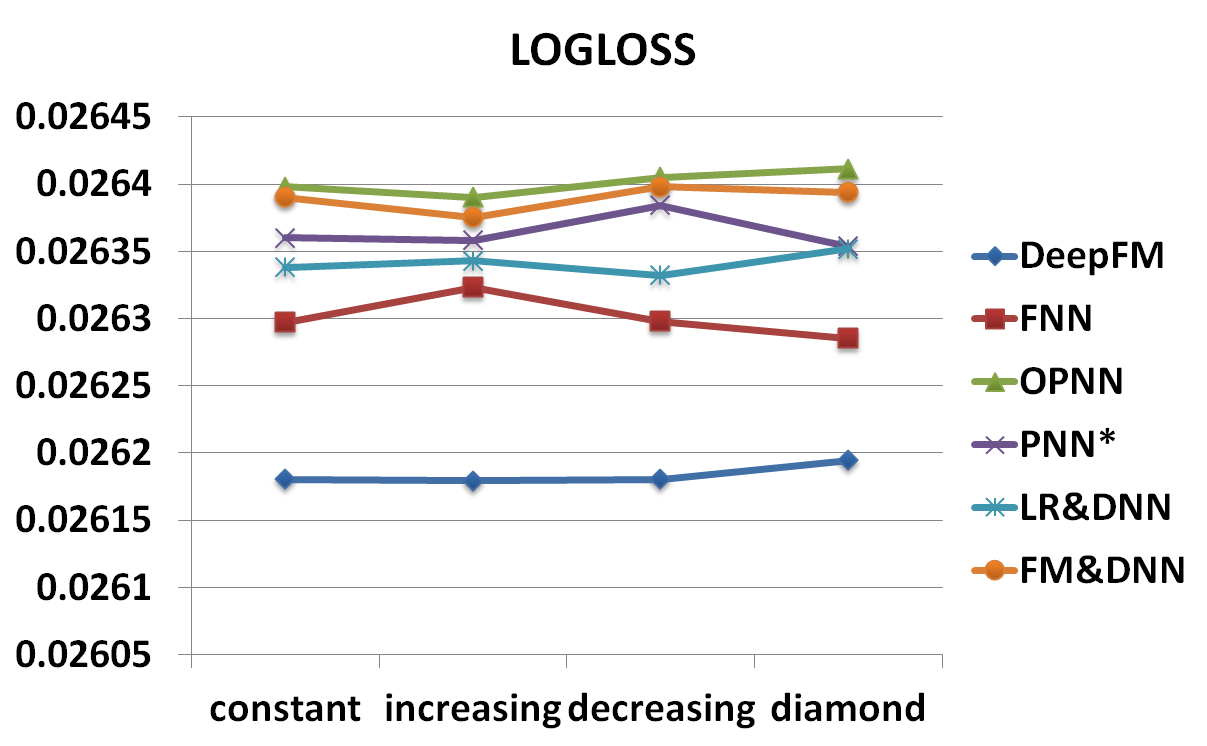}
\end{minipage}
\caption{\footnotesize{AUC and Logloss comparison of network shape.}}\label{fig:shape}
\end{figure}

We test four different network shapes: constant, increasing, decreasing, and diamond. When we change the network shape, we fix the number of hidden layers and the total number of neurons. For instance, when the number of hidden layers is 3 and the total number of neurons is 600, then four different shapes are: constant (200-200-200), increasing (100-200-300), decreasing (300-200-100), and diamond (150-300-150). As we can see from Figure~\ref{fig:shape}, the ``constant" network shape is empirically better than the other three options, which is consistent with previous studies~\cite{networkstructure09}.

%% file: related.tex
\section{Related Work}\label{section:related}

In this paper, a new deep neural network is proposed for CTR prediction. The most related domains are CTR prediction and deep learning in recommender system. In this section, we discuss related work in these two domains.

CTR prediction plays an important role in recommender system~\cite{CTR07,ffm,FTRL}. Besides generalized linear models and FM, a few other models are proposed for CTR prediction, such as tree-based model~\cite{facebookgbdt}, tensor based model~\cite{PITFRendleS10}, support vector machine~\cite{Poly2SVM10}, and bayesian model~\cite{bayesCTR10}.

%Among these models, generalized linear models such as \cite{CTR07,FTRL} are widely used in industry applications because of their simplicity and scalability, however, they are hard to capture high-order feature interactions automatically. For the purpose of capturing order-2 feature interactions automatically, factorization machine approaches such as \cite{fm,ffm} are deployed in real application. Moreover, another non-linear tree-based model~\cite{facebookgbdt} is used for capture non-linear features.

The other related domain is deep learning in recommender systems. In Section~\ref{section:intro} and Section~\ref{section:App:rela}, several deep learning models for CTR prediction are already mentioned, thus we do not discuss about them here. Several deep learning models are proposed in recommendation tasks other than CTR prediction (e.g., \cite{youtube,other-2,other-7,other-6,other-5,wsdm17recurentRS,jointdeepforreviewZhengNY17}). \cite{other-2,other-3,other-4} propose to improve Collaborative Filtering via deep learning. The authors of \cite{other-1,other-7} extract content feature by deep learning to improve the performance of music recommendation. \cite{cnn_image} devises a deep learning network to consider both image feature and basic feature of display adverting. \cite{youtube} develops a two-stage deep learning framework for YouTube video recommendation.

%% file: conclusion.tex
\section{Conclusions}\label{section:conclusion}

In this paper, we proposed DeepFM, a Factorization-Machine based Neural Network for CTR prediction, to overcome the shortcomings of the state-of-the-art models and to achieve better performance. DeepFM trains a deep component and an FM component jointly. It gains performance improvement from these advantages: 1) it does not need any pre-training; 2) it learns both high- and low-order feature interactions; 3) it introduces a sharing strategy of feature embedding to avoid feature engineering. We conducted extensive experiments on two real-world datasets (Criteo dataset and a commercial App Store dataset) to compare the effectiveness and efficiency of DeepFM and the state-of-the-art models. Our experiment results demonstrate that 1) DeepFM outperforms the state-of-the-art models in terms of AUC and Logloss on both datasets; 2) The efficiency of DeepFM is comparable to the most efficient deep model in the state-of-the-art.

There are two interesting directions for future study. One is exploring some strategies (such as introducing pooling layers) to strengthen the ability of learning most useful high-order feature interactions. The other is to train DeepFM on a GPU cluster for large-scale problems.

%% file: paper.bbl
\begin{thebibliography}{}

\bibitem[\protect\citeauthoryear{Boulanger{-}Lewandowski \bgroup \em et
  al.\egroup }{2013}]{audioBoulanger-LewandowskiBV13}
Nicolas Boulanger{-}Lewandowski, Yoshua Bengio, and Pascal Vincent.
\newblock Audio chord recognition with recurrent neural networks.
\newblock In {\em ISMIR}, pages 335--340, 2013.

\bibitem[\protect\citeauthoryear{Chang \bgroup \em et al.\egroup
  }{2010}]{Poly2SVM10}
Yin{-}Wen Chang, Cho{-}Jui Hsieh, Kai{-}Wei Chang, Michael Ringgaard, and
  Chih{-}Jen Lin.
\newblock Training and testing low-degree polynomial data mappings via linear
  {SVM}.
\newblock {\em JMLR}, 11:1471--1490, 2010.

\bibitem[\protect\citeauthoryear{Chen \bgroup \em et al.\egroup
  }{2016}]{cnn_image}
Junxuan Chen, Baigui Sun, Hao Li, Hongtao Lu, and Xian{-}Sheng Hua.
\newblock Deep {CTR} prediction in display advertising.
\newblock In {\em MM}, 2016.

\bibitem[\protect\citeauthoryear{Cheng \bgroup \em et al.\egroup
  }{2016}]{wide-n-deep}
Heng{-}Tze Cheng, Levent Koc, Jeremiah Harmsen, Tal Shaked, Tushar Chandra,
  Hrishi Aradhye, Glen Anderson, Greg Corrado, Wei Chai, Mustafa Ispir, Rohan
  Anil, Zakaria Haque, Lichan Hong, Vihan Jain, Xiaobing Liu, and Hemal Shah.
\newblock Wide {\&} deep learning for recommender systems.
\newblock {\em CoRR}, abs/1606.07792, 2016.

\bibitem[\protect\citeauthoryear{Covington \bgroup \em et al.\egroup
  }{2016}]{youtube}
Paul Covington, Jay Adams, and Emre Sargin.
\newblock Deep neural networks for youtube recommendations.
\newblock In {\em RecSys}, pages 191--198, 2016.

\bibitem[\protect\citeauthoryear{Graepel \bgroup \em et al.\egroup
  }{2010}]{bayesCTR10}
Thore Graepel, Joaquin~Qui{\~{n}}onero Candela, Thomas Borchert, and Ralf
  Herbrich.
\newblock Web-scale bayesian click-through rate prediction for sponsored search
  advertising in microsoft's bing search engine.
\newblock In {\em ICML}, pages 13--20, 2010.

\bibitem[\protect\citeauthoryear{He \bgroup \em et al.\egroup
  }{2014}]{facebookgbdt}
Xinran He, Junfeng Pan, Ou~Jin, Tianbing Xu, Bo~Liu, Tao Xu, Yanxin Shi,
  Antoine Atallah, Ralf Herbrich, Stuart Bowers, and Joaquin~Qui{\~{n}}onero
  Candela.
\newblock Practical lessons from predicting clicks on ads at facebook.
\newblock In {\em ADKDD}, pages 5:1--5:9, 2014.

\bibitem[\protect\citeauthoryear{He \bgroup \em et al.\egroup
  }{2016}]{residual2016}
Kaiming He, Xiangyu Zhang, Shaoqing Ren, and Jian Sun.
\newblock Deep residual learning for image recognition.
\newblock In {\em CVPR}, pages 770--778, 2016.

\bibitem[\protect\citeauthoryear{Juan \bgroup \em et al.\egroup }{2016}]{ffm}
Yu{-}Chin Juan, Yong Zhuang, Wei{-}Sheng Chin, and Chih{-}Jen Lin.
\newblock Field-aware factorization machines for {CTR} prediction.
\newblock In {\em RecSys}, pages 43--50, 2016.

\bibitem[\protect\citeauthoryear{Larochelle \bgroup \em et al.\egroup
  }{2009}]{networkstructure09}
Hugo Larochelle, Yoshua Bengio, J{\'{e}}r{\^{o}}me Louradour, and Pascal
  Lamblin.
\newblock Exploring strategies for training deep neural networks.
\newblock {\em JMLR}, 10:1--40, 2009.

\bibitem[\protect\citeauthoryear{Liu \bgroup \em et al.\egroup }{2015}]{cnn}
Qiang Liu, Feng Yu, Shu Wu, and Liang Wang.
\newblock A convolutional click prediction model.
\newblock In {\em CIKM}, 2015.

\bibitem[\protect\citeauthoryear{McMahan \bgroup \em et al.\egroup
  }{2013}]{FTRL}
H.~Brendan McMahan, Gary Holt, David Sculley, Michael Young, Dietmar Ebner,
  Julian Grady, Lan Nie, Todd Phillips, Eugene Davydov, Daniel Golovin, Sharat
  Chikkerur, Dan Liu, Martin Wattenberg, Arnar~Mar Hrafnkelsson, Tom Boulos,
  and Jeremy Kubica.
\newblock Ad click prediction: a view from the trenches.
\newblock In {\em KDD}, 2013.

\bibitem[\protect\citeauthoryear{Qu \bgroup \em et al.\egroup }{2016}]{pnn}
Yanru Qu, Han Cai, Kan Ren, Weinan Zhang, Yong Yu, Ying Wen, and Jun Wang.
\newblock Product-based neural networks for user response prediction.
\newblock {\em CoRR}, abs/1611.00144, 2016.

\bibitem[\protect\citeauthoryear{Rendle and
  Schmidt{-}Thieme}{2010}]{PITFRendleS10}
Steffen Rendle and Lars Schmidt{-}Thieme.
\newblock Pairwise interaction tensor factorization for personalized tag
  recommendation.
\newblock In {\em WSDM}, pages 81--90, 2010.

\bibitem[\protect\citeauthoryear{Rendle}{2010}]{fm}
Steffen Rendle.
\newblock Factorization machines.
\newblock In {\em ICDM}, 2010.

\bibitem[\protect\citeauthoryear{Richardson \bgroup \em et al.\egroup
  }{2007}]{CTR07}
Matthew Richardson, Ewa Dominowska, and Robert Ragno.
\newblock Predicting clicks: estimating the click-through rate for new ads.
\newblock In {\em WWW}, pages 521--530, 2007.

\bibitem[\protect\citeauthoryear{Salakhutdinov \bgroup \em et al.\egroup
  }{2007}]{other-2}
Ruslan Salakhutdinov, Andriy Mnih, and Geoffrey~E. Hinton.
\newblock Restricted boltzmann machines for collaborative filtering.
\newblock In {\em ICML}, pages 791--798, 2007.

\bibitem[\protect\citeauthoryear{Sedhain \bgroup \em et al.\egroup
  }{2015}]{other-3}
Suvash Sedhain, Aditya~Krishna Menon, Scott Sanner, and Lexing Xie.
\newblock Autorec: Autoencoders meet collaborative filtering.
\newblock In {\em WWW}, pages 111--112, 2015.

\bibitem[\protect\citeauthoryear{Srivastava \bgroup \em et al.\egroup
  }{2014}]{dropout14}
Nitish Srivastava, Geoffrey~E. Hinton, Alex Krizhevsky, Ilya Sutskever, and
  Ruslan Salakhutdinov.
\newblock Dropout: a simple way to prevent neural networks from overfitting.
\newblock {\em JMLR}, 15(1):1929--1958, 2014.

\bibitem[\protect\citeauthoryear{van~den Oord \bgroup \em et al.\egroup
  }{2013}]{other-7}
A{\"{a}}ron van~den Oord, Sander Dieleman, and Benjamin Schrauwen.
\newblock Deep content-based music recommendation.
\newblock In {\em NIPS}, pages 2643--2651, 2013.

\bibitem[\protect\citeauthoryear{Wang and Wang}{2014}]{other-1}
Xinxi Wang and Ye~Wang.
\newblock Improving content-based and hybrid music recommendation using deep
  learning.
\newblock In {\em {ACM} {MM}}, pages 627--636, 2014.

\bibitem[\protect\citeauthoryear{Wang \bgroup \em et al.\egroup
  }{2015}]{other-4}
Hao Wang, Naiyan Wang, and Dit{-}Yan Yeung.
\newblock Collaborative deep learning for recommender systems.
\newblock In {\em {ACM} {SIGKDD}}, pages 1235--1244, 2015.

\bibitem[\protect\citeauthoryear{Wu \bgroup \em et al.\egroup }{2016}]{other-6}
Yao Wu, Christopher DuBois, Alice~X. Zheng, and Martin Ester.
\newblock Collaborative denoising auto-encoders for top-n recommender systems.
\newblock In {\em {ACM} {WSDM}}, pages 153--162, 2016.

\bibitem[\protect\citeauthoryear{Wu \bgroup \em et al.\egroup
  }{2017}]{wsdm17recurentRS}
Chao{-}Yuan Wu, Amr Ahmed, Alex Beutel, Alexander~J. Smola, and How Jing.
\newblock Recurrent recommender networks.
\newblock In {\em WSDM}, pages 495--503, 2017.

\bibitem[\protect\citeauthoryear{Zhang \bgroup \em et al.\egroup }{2014}]{rnn}
Yuyu Zhang, Hanjun Dai, Chang Xu, Jun Feng, Taifeng Wang, Jiang Bian, Bin Wang,
  and Tie{-}Yan Liu.
\newblock Sequential click prediction for sponsored search with recurrent
  neural networks.
\newblock In {\em AAAI}, 2014.

\bibitem[\protect\citeauthoryear{Zhang \bgroup \em et al.\egroup }{2016}]{fnn}
Weinan Zhang, Tianming Du, and Jun Wang.
\newblock Deep learning over multi-field categorical data - - {A} case study on
  user response prediction.
\newblock In {\em ECIR}, 2016.

\bibitem[\protect\citeauthoryear{Zheng \bgroup \em et al.\egroup
  }{2016}]{other-5}
Yin Zheng, Yu{-}Jin Zhang, and Hugo Larochelle.
\newblock A deep and autoregressive approach for topic modeling of multimodal
  data.
\newblock {\em {IEEE} Trans. Pattern Anal. Mach. Intell.}, 38(6):1056--1069,
  2016.

\bibitem[\protect\citeauthoryear{Zheng \bgroup \em et al.\egroup
  }{2017}]{jointdeepforreviewZhengNY17}
Lei Zheng, Vahid Noroozi, and Philip~S. Yu.
\newblock Joint deep modeling of users and items using reviews for
  recommendation.
\newblock In {\em WSDM}, pages 425--434, 2017.

\end{thebibliography}
